\documentclass[a4paper,12pt]{article}

\usepackage[latin1]{inputenc}
\usepackage[american]{babel}
\usepackage[dvips]{graphicx}
\usepackage{amsmath} \usepackage{amsfonts} \usepackage{amssymb}
\usepackage{bbm}

\numberwithin{equation}{section}

\newcommand{\D}{{\bf \not}  D}
\newcommand{\nn}{\nonumber}
\newcommand{\tr}{\mbox{tr}}    

\begin{document}

\hyphenation{fer-mi-ons stren-ghts}

\begin{flushright}
IC/2006/130\\
SISSA-80/2006/EP

\end{flushright}
\begin{flushright}
\end{flushright}

\begin{center}
\huge{\sc Warped Models in String Theory}\\
\vskip 2cm

\Large{\sc B.S. Acharya$^a$, F. Benini$^b$ 
 and
R. Valandro$^b$
}
\\
\vskip 8mm \normalsize
{\sf $^a$ Abdus Salam ICTP,
Strada Costiera 11,\\ 34014 Trieste, ITALIA\\
\smallskip
and \\
\smallskip
INFN, sezione di Trieste\\
\smallskip
{\it bacharya\_at\_cern.ch}\\
\vskip 5mm
$^b$ International School for Advanced Studies (SISSA/ISAS),\\Via Beirut 2-4, 34014 Trieste, ITALIA \\
\smallskip
{\it benini, valandro\_at\_sissa.it}}

\end{center}
\vskip 3cm
\begin{center}
{\bf {\sc Abstract}}
\end{center}

Warped models, originating with the ideas of Randall and Sundrum, provide a fascinating extension
of the standard model with interesting consequences for the LHC. We investigate in detail how string theory
realises such models, with emphasis on fermion localisation and the computation of Yukawa couplings. We find,
in contrast to the 5d models, that fermions can be localised {\it anywhere} in the extra dimension, and that there are {\it new mechanisms} to generate exponential hierarchies amongst the Yukawa couplings. We also suggest a way to distinguish these string theory models with data from the LHC.

\newpage

\section{Introduction}

Since the seminal papers of Randall and Sundrum \cite{rs}, models of particle physics constructed in
a five dimensional warped geometry have offered insights into the hierarchy problem, fermion masses
and many other issues related to the standard model.
Since the original ideas proposed in \cite{rs}, the state of the art 5d models have evolved
somewhat \cite{5d1,5d2,5d3,5d4,5d5,5d6,5d7,5d8,5d9,5d10,5d11,5d12,5d13,5d14,5d15,5d16,5d17} 
(see, for instance, \cite{ghergh1} for a review). Moreover, there are potentially very interesting signals 
for the LHC, since these models are
dual descriptions of `compositeness' \cite{4ddual}.
Their most basic features are:
\begin{itemize}
\item[a)] for every standard model field, there is a bulk 5d field;
\item[b)] to solve the hierarchy problem, the Higgs is localised in a region of large warping; 
\item[c)] turning on bulk and boundary masses localises the fermion zero modes and hence one obtains hierarchical Yukawa couplings since the fermions can have varying degrees of overlap with the Higgs.
\end{itemize}
Since these models have arbitrary parameters {\it e.g.} the bulk and boundary masses, we would like to investigate
the realisation of these models in string theory. This perspective offers a framework for explaining the parameters of the 5d models and some new insights:
\begin{itemize}
\item To realise a warped geometry we consider warped string compactifications which arise naturally in the IIB string theory with fluxes \cite{GKP,verlinde}.
\item Matter and gauge fields in the bulk arise as strings which end on $D$7-branes in the bulk\footnote{
Previous studies of warped models in string theory have tended to have the standard model on $D$3-branes
\cite{uranga} See however \cite{gherghetta}.}.
\item To have several standard model generations, we turn on a topologically non-trivial (``instanton'') background field on the $D$7 worldvolume. 
\item Fermion zero modes then naturally localise near the instantons and/or by warping.
This leads to new features: a) the zero modes can be {\it localised anywhere} in the extra dimension and b) the scale of the topologically non-trivial background (instanton size) can also be used to suppress Yukawa couplings, in addition to the usual mechanism of
separating the fermion zero modes in the extra dimension.
\end{itemize}

Our main results will be explicit formulae for the profile of the fermion zero modes in the 5th dimension and their Yukawa couplings. 
These formulae show in particular how the
physical
size of the topologically non-trivial ``instanton'' background field can give rise to
hierarchies of Yukawa couplings. They also show that the large Yukawa coupling is associated
with a ``small instanton'' in the extra dimension.
In the next section, which is relatively self-contained, we will illustrate 
these results in the simplest possible example.
Following this we go on to describe more general results
including examples in which there can be several ``instantons'' in the background, 
leading to new effects. We conlude the paper with a discussion of the differences between the
string theory vacua and the 5d models and suggest a way in which the LHC data could be used to
distinguish amongst them.
The appendices
deal with technicalities on the fermion zero modes and the Dirac operator.

\newpage

\section{A Simple Example}

In this section we will describe a simple example which illustrates most of the features of the more
general calculations that are given in the rest of the paper.

Our interest is understanding how various features of the 5d phenomenological models are realised in string
theory vacua, with the motivation that this might lead to additional insights about the phenomenology. The three
basic features which we wish to understand better are:
\begin{itemize}
\item[a)] The 5d warped models tend to have the standard model gauge fields propagating in the bulk of $AdS_5$.
\item[b)] For each standard model fermion there is a 5d bulk fermion field with both bulk and boundary mass parameters which determine whether or not the fermion is localised in the UV or IR end of $AdS_5$.
\item[c)] The hierarchy amongst standard model Yukawa couplings is realised by the varying degrees of overlap
between these localised wavefunctions and the Higgs.
\end{itemize}
We will study the string theory realisation of these features within the context of Type IIB 
string theory vacua with fluxes, since this class of vacua realises warped extra dimensions in a natural
way. In such vacua, non-Abelian gauge fields can reside on $D$3 and $D$7-branes, so in order to realise
property a) the only possibility is to put the standard model gauge fields on the $D$7-branes. Recall that
the ten dimensional spacetime is a warped product of four-dimensional Minkowski spacetime $M^{3,1}$ and
a compact Calabi-Yau manifold $Z$ \cite{GKP}. The metric takes the form of a $D$3-brane metric, where the $D$3-branes span the Minkowski spacetime.
The $D$7-branes have a world-volume which is a warped product of
$M^{3,1}$ and a four dimensional cycle $\Sigma \subset Z$.

Now we turn to property b). The physics behind the introduction of bulk and boundary masses is that, before
symmetry breaking, the standard model fermions are all
zero modes of the Dirac operator on $M^{3,1}$. We thus need to study the Dirac equation on
the $D$7-brane in the
warped background. For the ten dimensional geometries described in \cite{GKP} the metric induced on
the $D$7-branes is of the form:
\begin{equation}
ds_8^2 = f(z)^{-1/2} \eta_{\mu\nu} dx^\mu dx^\nu + f(z)^{1/2} \,g_{\alpha\beta}\,dz^\alpha dz^\beta \qquad (\alpha,\beta=1,...,4) \:,
\end{equation}
where the warp factor $f$ is a function of the coordinates $z^\alpha$ on the 4-cycle $\Sigma$, which 
the $D$7-brane wraps
and $x^{\mu}$ are coordinates on $M^{3,1}$. For simplicity, we will study the
warped geometry induced by $D$3-branes in flat spacetime. Most of the interesting features we observe are not very sensitive to the geometry of $\Sigma$ as we will see.
In this case
\begin{equation}
ds_8^2 = f(r)^{-1/2} \eta_{\mu\nu} dx^\mu dx^\nu + f(r)^{1/2} \,\delta_{\alpha\beta}\,dz^\alpha dz^\beta \qquad (\alpha,\beta=1,...,4) \:,
\end{equation}
where $f(r) = 1+L^4/(r^2+d_0^2)^2$, $r^2= |\vec z|^2$ and $d_0$ is the separation between the $D$7 and the $D$3-branes. For simplicity, in this example, we will set $d_0=0$.

We will also use an almost ``flat'' radial coordinate $y$ defined by
\begin{equation}
r = L \, e^{-ky} \qquad \qquad k=\frac{1}{L}
\end{equation} 
For illustration, the near horizon geometry in these coordinates is
\begin{gather}
ds_8^2 = e^{-2ky} \eta_{\mu\nu} dx^\mu dx^\nu + dy^2  + L^2 d\Omega_3^2 \\
\end{gather}
which is an $AdS_5 \times S^3$ contained in $AdS_5 \times S^5$.
In these coordinates, $y\to\infty$ is the tip of the throat while $y=0$ is its origin.
 
The low energy spectrum of the $D$7-brane modes includes massless fermions in the adjoint representation of the gauge group:
$\D_8 \Psi = 0$. Under the splitting induced by the $D$3-brane background, the fermions factorise as products
of fermions on $M^{3,1}$ and $\Sigma = \mathbb{R}^4$:
\begin{equation} \label{fermions splitting}
\Psi = \sum\nolimits_k \chi_k(x) \otimes d_{\psi_k} \, \psi_k(z)  \;,
\end{equation}
where $d_{\psi_k}$ is a normalization constant.

The Dirac equation can be written as\footnote{In our conventions, $\Gamma^\mu$ are the gamma matrices relative to the background metric, while $\gamma^\mu$ are relative to the flat metric.} 
\begin{equation}
\D_8 \Psi = \biggl( f^{1/4}\tilde{\D}_{3,1} + \frac{1}{f^{1/4}}\tilde{\D}_4 - \frac{1}{8f^{1/4}} \frac{f'}{f} \, 
		\gamma_r \biggr) \, \Psi =0\;,
\end{equation}
where $\tilde{\D}_{3,1}$ and $\tilde{\D}_4$ are respectively the Dirac operator on $M^{3,1}$ and on flat $\mathbb{R}^4$. Massless fermions in $M^{3,1}$ are the zero modes of $\bigl( \tilde{\D}_4 - \frac{f'}{8f} \, \gamma_r \bigr)$. As shown in the appendix these are given by: 
\begin{equation}\label{eqRfermzeromd}
\psi = f^{1/8} \tilde{\psi} \:, \nn
\end{equation}
where $\tilde{\psi}$ are the zero modes of the operator $\tilde{\D}_4$. This means that in the warped background, the {\it four-dimensional zero modes are conformally equal to the zero modes in an unwarped geometry}.

\

The simplest possibility in this example is to take $\tilde{\psi}$ to be the constant zero modes
of the flat Euclidean Dirac operator $\tilde{\D}_4$ on the extra dimensions. Whilst this indeed will give us
a four dimensional fermion zero mode, it raises two problems:
\begin{enumerate}
\item since the fermion field $\Psi$ on the
$D$7-brane is in the adjoint representation, the four dimensional zero mode $\psi$ is also in the adjoint
representation;
\item there will be  {\it four} such fermion zero modes (since there are four constant spinors), whilst the standard model requires three generations of zero modes in representations which are certainly not adjoint.
\end{enumerate}
In principle, there is an elegant
solution to both of these problems, which also elucidates the string theory description of property $b)$:
the gauge covariant Dirac operator $\tilde{\D}_4$ can have multiple non-trivial zero modes in the presence of topologically non-trivial gauge
field backgrounds. This is a standard mechanism to generate light fermion generations in string theory, however the novelty here is the presence of the warp factor in $\psi$ and that we will be quite explicit about
the profile of the wavefunction.

As is well known from gauge theory instanton physics, gauge field-strengths satisfying the condition $F=-\ast_4 F$ in four Euclidean dimensions
(usually called instantons) can be topologically non-trivial and support multiple
fermion zero modes which are not in the adjoint representation. Depending on the topological charge
(or instanton number) one can have different numbers of fermion zero modes.
One can check that such gauge field configurations also solve the equations of motion on the $D$7-brane, so are acceptable backgrounds.

The zero mode wave functions $\tilde{\psi}$ have been computed explicitly long ago 
for many different $F=-\ast_4 F$
backgrounds \cite{zeros}. 
If we take the simplest known solution to these equations \cite{tHooft}, then we obtain
a zero mode which depends on the {\it size} of the instanton $\rho$,  as well as its position $\vec Z_\psi$
in the Euclidean space: 
\begin{equation}
\psi(\vec z)= f^{1/8} \frac{\rho}{\big[ \rho^2+( \vec z - \vec Z_\psi)^2 \big]^{3/2}} \; \eta \;,
\end{equation}
here $\eta$ is a constant spinor normalized as $\eta^\dag \eta=1$.

These fermion zero modes have to be normalised properly. Consider the kinetic term:
\begin{multline}
-\int d^8x \sqrt{-G}\: G^{\mu\nu} \: \bar{\Psi}\Gamma_\mu \partial_\nu \Psi + ...\\
= -\int d^4x \, \eta^{\mu\nu}\,\bar{\chi}(x) \gamma_\mu \partial_\nu \chi(x)
\int d^4z \: d_\psi^2 \,f^{1/4} (z) \: \psi(z)^\dagger \psi(z) + ...
\end{multline}
where the normalisation constant $d_{\psi}$ was introduced in the Kaluza-Klein ansatz \eqref{fermions splitting}
and we used $\Gamma_\mu = f^{-1/4}\gamma_\mu$.
In order to have a canonical kinetic term, we require:
\begin{eqnarray}
 d_\psi^2 \int d^4z\, f^{1/4}\,\psi^\dagger \psi&=& 1
\end{eqnarray}
In regions of negligible warping, this condition is realised for $d_\psi\sim 1$, whilst when the warp factor is large (for instance in the near horizon region) the normalisation is given by:
\begin{align}
d_\psi^{-2} &= \int r^3dr \, 4\pi\sin\theta \, d\theta \, f(r)^{1/2}\, \frac{\rho^2}{(\rho^2+r^2 +Z_\psi^2-2r Z_\psi\cos\theta)^{3}} \\
&= \frac{\pi^2}{2}\left(\frac{\rho^2}{L^2}+e^{-2kY_\psi} \right)^{-1} \:,
\end{align}
where $|\vec Z_\psi|/L \equiv e^{-kY_\psi}$ is the radial position of the instanton in almost flat radial coordinates.
When $\rho/L < e^{-kY_\psi}$, we get $d_\psi \simeq (\sqrt{2}/\pi) \, e^{-k Y_\psi}$.

We see that in string theory
the instanton scale size is important in determining the profile of the fermion zero modes. Putting
all the factors together, the {\it normalised} zero mode wave function is:
\begin{equation}
d_\psi \, \psi \sim e^{-k Y_\psi} e^{\frac{k}{2}y} \frac{\rho}{\Bigl[ \rho^2 + (\vec z - \vec Z_\psi)^2 \Bigr]^{3/2}} \: \eta
\end{equation}
We can compare this wave function with the 5d models:
\begin{equation}
d_{\psi 5d} \, \psi_{5d} \sim \sqrt{\frac{k (1-2c)}{e^{(1-2c)kR} -1}} e^{(2-c)ky} \;,
\end{equation}
where the constant $c$ is a combination of bulk and boundary masses. 

From this we learn that the zero mode wavefunction in string theory is quite different from the
5d models. Note that there is a dependence on the instanton scale size, $\rho$. 
In particular, in string theory {\it the zero mode can be localised anywhere}
in the 5th dimension. 

\subsection{Instantons as $D$3-branes}

As is well known, gauge field backgrounds on $D$7-branes with $F\wedge F \neq 0$ carry $D$3-brane
charge \cite{mrd}. 
In fact, smooth instanton backgrounds such as those we are considering here, are ``fat $D$3-branes''
with size $\rho$. Therefore, we can also say that the fermion zero modes are localised on fat $D$3-branes. 
The fermion zero modes are therefore 3-7 strings.
Note however that, in order to trust the metric we have been using, we should consider the number of such fat $D$3-branes to be small compared to the large number of ordinary $D$3-branes
and fluxes which generate the bulk geometry.

The parameters $\rho$ and $\vec Z_{\psi}$ are therefore moduli field vevs which arise in the open string
sector. It would be interesting to investigate mechanisms which stabilise these moduli. Presumably
closed and open string fluxes generate a potential for these fields. 

\

\subsection{Yukawa Couplings}

The zero mode profiles are
crucial for computing the 4d Yukawa couplings, and clearly the answer will depend on $\rho$. In order to
determine the Yukawa couplings, we need to identify the Higgs field in string theory. Essentially, with
only $D$3 and $D$7-branes the Higgs can be a 7-7 or a 3-7 string, since it must be charged under the standard model gauge group. The simplest case to consider is
that the Higgs is a 3-7 string state. The 7-7 case will be described later.
In this case its wavefunction will be localised near a point $\vec Z_H$ in $\Sigma$ and we  will simply model this by a delta-function.
This choice is very similar to the standard 5d proposal \cite{ghergh1}.

We must first determine the correctly normalised 4d Higgs field from its kinetic term by imposing
\begin{multline}
-\int d^8x \sqrt{-\hat G_{3,1}}\: G^{\mu\nu} \: d_H^2 \: \partial_\mu \bar{H}(x)\partial_\nu H(x) \, \delta(\vec z - \vec Z_H) = \\
= - \int d^4x \, \partial_\mu \bar{H}(x)\partial^\mu H(x)
\end{multline}
which gives $d_H= f(|\vec Z_H|)^{1/4}$.

The four dimensional Yukawa coupling is obtained by direct dimensional reduction of the 8d one (remembering localisation of the Higgs):
\begin{multline}
\int d^8x \, \sqrt{-\hat G_{3,1}} \, \lambda^{(8)} \, d_H \, \bar \Psi \Psi H \, \delta(\vec z - \vec Z_H) = \\
= \lambda^{(8)} d_H d_\psi^2 f(|\vec Z_H|)^{-1} \psi(\vec Z_H)^2 \int d^4x \, \bar \chi(x) \chi(x) H(x) \:,
\end{multline}
so that
\begin{equation}
\lambda = \lambda^{(8)} d_\psi^2 \, \left. \frac{\psi^2(z)} {f(z)^{3/4}} \right|_{\vec Z_H} \:.
\end{equation} 
Remember that the 8d Yukawa has dimension of (length)$^4$. We see therefore that the Yukawa
coupling in the standard model is determined by several factors: the fermion zero mode evaluated at
the Higgs position, the warp factor at the Higgs position and the normalisation constant $d_{\psi}$
(which itself depends on $\rho$ and $Y_{\psi}$).

\

Let us analyse the 4d Yukawa coupling further. For simplicity we will study the case when
the fermion zero mode is localised in a region of large warping and  $\rho/L < e^{-kY_\psi}$. Then the 4d Yukawa coupling is given by:
\begin{equation}\label{Yukawa}
\lambda = \frac{2}{\pi^2} \lambda^{(8)} e^{-2k(Y_H + Y_\psi)} \frac{\rho^2}{\big[ \rho^2 + (\vec Z_H - \vec Z_\psi)^2 \big]^3} \:, \nonumber
\end{equation}
where again we used almost flat radial coordinates $|\vec Z_H|/L \equiv e^{-kY_H}$. In the standard model
the Yukawa couplings of the charged fermions range from order one for the top quark to $10^{-6}$ for the
electron, and clearly (\ref{Yukawa}) is rich enough to span this range. In more detail,
the top quark Yukawa coupling ($\lambda\sim 1$) can arise when the top wave function peaks at the
location of the Higgs {\it i.e.}  $Y_H = Y_\psi$:
\begin{equation}\label{eqR019}
\lambda = \frac{2}{\pi^2} \frac{\lambda^{(8)}}{\rho^4} e^{-4kY_H}
\end{equation}
Notice that, due to the warping in the spacetime,
$\rho$ is not the physical size $\rho_{phys}$ of the instanton, which depends upon its location in $AdS_5$:
\begin{equation}\label{physsize}
\rho_\text{phys} = \int_{|\vec Z_\psi|-\frac{\rho}{2}}^{|\vec Z_\psi|+\frac{\rho}{2}} ds = \int_{|\vec Z_\psi|-\frac{\rho}{2}}^{|\vec Z_\psi|+\frac{\rho}{2}} f^{1/4}(r) dr \simeq e^{kY_\psi} \rho \:,
\end{equation}
where the last result is valid when $\rho < Le^{-kY_\psi}$. The same can be seen by evaluating the instanton displacement in the almost flat radial coordinate: $\Delta y = e^{kY_\psi} \rho$. 
Note that in terms of the physical size, this is simply $\rho_\text{phys} < L$: 
the instanton is physically smaller than the $AdS_5$ radius, which is a natural requirement.
Substituting in \eqref{eqR019}, one gets:
\begin{equation}
\lambda = \frac{2}{\pi^2} \frac{\lambda^{(8)}}{\rho_{phys}^4}\:.
\end{equation}
In general, we expect $\lambda^{(8)}$ to be of order $\ell^4$, with $\ell$ the string scale, we obtain $\lambda \sim 1$ when $\rho_\text{phys}$ is of order of the string scale. In other words, {\it the instanton which localises the top quark is a small instanton.} We therefore might expect strong quantum corrections to the top sector.
On the other hand, when $\rho_\text{phys}$ is larger than the fundamental scale, $\lambda$ is smaller than 1 and we can also realize smaller Yukawa couplings by localising the corresponding fermions on large
instantons.

\

The smaller Yukawa couplings are actually better obtained in the case when $Y_\psi < Y_H$, which means that the fermion zero mode is localised far from the Higgs, and again when $\rho < L e^{-kY_\psi}$. The Yukawa coupling is then given by:
\begin{equation}
\lambda = \frac{2}{\pi^2} \lambda^{(8)} \, e^{-2k(Y_H -2Y_\psi)} \, \frac{\rho^2}{L^6} \:.
\end{equation}
This can be written as 
\begin{equation}
\lambda = \frac{2}{\pi^2} \, \frac{\lambda^{(8)}}{\rho_\text{phys}^4} \, \frac{\rho_\text{phys}^6}{L^6} \, e^{-2k(Y_H-Y_\psi)} \:.
\end{equation}
We see that even when the $AdS_5$ radius $L$ is just a little bigger than the instanton size, 
that {\it both the instanton scale size and the warp factor suppress the generic Yukawa coupling.}


\section{The Higgs as a Vector Zero Mode}

In this section we will study the case that the Higgs is a 7-7 string which is a zero mode
of the 8-dimensional gauge field on the $D$7-brane. 
We will see that such zero modes are not affected by the presence of the warping and can also be computed in the instanton background. 

In the 8-dimensional kinetic term, all the fields are in the adjoint representation of the gauge group $G$. The background instanton gauge field breaks this group, leaving a (3+1)-dimensional gauge theory, whose gauge group is a subgroup $G'$ of $G$. The adjoint representation of $G$ splits into irreps of $G'\times SU(2)$, where $SU(2)$ is chosen as the gauge group of the instanton. Thus, an 8-dimensional field in the Adj rep of $G$ can be written as a sum of products of fields in $M^{3,1}$ and $\Sigma$ in various representations
of $G' \times SU(2)$.
In order to reproduce a GUT theory at low energy, we could take $G'$ to contain some GUT group as a subgroup.

Let us see some details. The 8-dimensional kinetic term is:
\begin{eqnarray}
  &&\int d^8X \sqrt{-G} \bar{\Psi}\D \Psi 
\end{eqnarray}
and contains the term
\begin{eqnarray}
  g\int d^8X \sqrt{-G} \bar{\Psi}\not\!\! \delta \! A \Psi
  	&\supset& g\int d^4x \,\bar{\chi}_i(x) \chi_j(x) H_k(x)\,\int d^4y\, \psi_i^\dagger(y)\not\!\!\delta \! a_k(y) \psi_j(y)\nn
\end{eqnarray}
where $g$ is the 8d gauge coupling (of order $\ell^2$, with $\ell$ the string legth) and  where we have used the splitting \eqref{fermions splitting} of the fermion fields and that of the vector:
\begin{equation}
 A(x,y)_m dy^m =A_{bkg}(y)+\sum_k H^k(x) \delta a_k(y)\:.
\end{equation}
We see that the effective Yukawa coupling in $(3+1)$-dimensions is given by:
\begin{equation}\label{eqR008}
 g \, d_{\psi_i} d_{\psi_j} d_H \int d^4y \, \tilde{\psi}_i^\dagger(y) \not\!\!\delta \! a_k(y) \tilde{\psi}_j(y)\:.
\end{equation}
where we have substituted the expression \eqref{eqRfermzeromd} for the fermion zero modes $\psi$.
Note that the warp factor has disappeared; it only enters in the fermion normalisation constants\footnote{For this particular choice for the Higgs, its normalisation is not affected by the warping and will be put $d_H=1$}. The zero modes $\delta a_k(y)$ are warp factor independent
because the Yang-Mills action on $\Sigma$ is conformally invariant.

The fields $\psi_i$, $\psi_j$ and $\delta a_k$ are in the $SU(2)$ representations dictated by the splitting of Adj$G$ and by the $G'$ representations that one wants $H$, $\chi_i$ and $\chi_j$ to belong to.

We will compute the integral \eqref{eqR008} in the simple case in which the two fermions are in the fundamental representation of $SU(2)$, while the vector zero mode is in the adjoint. We will see that the coupling can be highly suppressed in the usual approximation of well separated instantons, and that this suppression is due to the localisation of the zero modes near individual single instantons. This justifies this simple choice of representations, since the localisation is characteristic of the zero modes in any representation. This is important, because the suppression works whatever $SU(2)$-representations are associated (by the splitting of Adj$G$) with the particular GUT-representations that one wants to find in the GUT Yukawa interaction terms.
It would be interesting to compute the integral exactly, since new phenomena might arise.

\

We will consider the 't Hooft solution with instanton number $k=2$. 
This solutions has $5k=10$ explicit parameters: $\rho_1$, $\rho_H$, $\vec Z_1$ and $\vec Z_H$. The zero mode profiles when $k>1$ are given in the appendix. We will also choose both the fermion zero modes in 
\eqref{eqR008} to be localised around $\vec Z_1$, while the vector one (the Higgs) is to be localised around $\vec Z_H$. We put $\vec Z_H$ in a region of large warping, in order to address the hierarchy problem.
We will see that, in order to have a sufficiently large top Yukawa coupling, one must have $\delta a$ sharply localised around $\vec Z_H$. 

We substitute the expressions \eqref{eqR014} and \eqref{eqR015} in \eqref{eqR008} and estimate it in several asymptotic regions of the parameter space of the $k=2$ solution. With more than one instanton, we will find a new suppression mechanism:  due to the localisation of wavefunctions at well separated points, suppression can also occur due to a hierarchy in the two instanton sizes $\rho_1$ and $\rho_H$. The maximal value of the integral is actually obtained when $|\vec Z_H-\vec Z_1|\ll \rho_1,\rho_H$ and $\rho_1\sim \rho_H$. 

Actually when  $|\vec Z_H-\vec Z_1|\ll \rho_1,\rho_H$, the parameter $X\equiv |\vec Z_1-\vec Z_H|$ disappears from the result, that is:
\begin{equation}\label{eqRYukVec}
 g \, d_\psi^2 \int  \psi_i^\dagger \sigma^\mu \delta \! A_\mu^\Phi \psi_j \simeq g \, d_\psi^2 \alpha^\Phi \int r^3 dr\,
	 \frac{\rho_1^2\rho_H^2}{(r^2+\rho_1^2+\rho_H^2)^4} = \, d_\psi^2 \frac{g \alpha^\Phi}{24} 
	\frac{\rho_1^2 \rho_H^2}{(\rho_1^2+\rho_H^2)^3}
\end{equation}
where $\delta \! A_\mu^\Phi$ is defined in \eqref{eqR015}, and where $\alpha^\Phi$ is a constant of order one.
The expression \eqref{eqRYukVec} takes its maximal value when $\rho_1\sim\rho_H$:
\begin{equation}\label{eqR016}
 g \, d_\psi^2 \int  \psi_i^\dagger \sigma^\mu \delta \! a_\mu^\Phi \psi_j \sim \, \frac{2}{\pi^2}\frac{g}{\rho_H^2} e^{-2\kappa Y_H} 
\end{equation}
The same result as \eqref{eqR016} is obtained taking $k=1$. 
Then one has to substitute the physical size in this formula (see \eqref{physsize}). The final result is:
\begin{eqnarray}
 \lambda &=& \frac{2}{\pi^2}\frac{g}{\rho_{H\text{phys}}^2}
\end{eqnarray}

From here, we see that if one wants the top coupling to be of order one, the top zero mode must be localised close to the Higgs and the $\rho$-parameter of the corresponding instanton has to be of the order of the Higgs one. 

\

The Yukawa hierarchy can then be obtained by varying the instanton parameters in such a way as
to have different overlaps of the zero modes. One can approximate the integral giving the Yukawa
couplings in different asymptotic regions of the instanton moduli space.
We summarize the results in Table \ref{tab}. In order to get the actual Yukawa coupling, this integral has to be multiplied by $d_\psi^2$ and the instanton `sizes' have to be substituted with their physical sizes. Let us consider some relevant cases, which turn out to be similar to the result found in the simple example of the previous section.
\begin{itemize}
\item When the fermions are localised around the same position of the Higgs:
	\begin{equation}
	\lambda = \frac{g}{\rho_{\psi\text{phys}}^2}\left( \frac{\rho_H}{\rho_\psi}\right)^2\nn
	\end{equation}
\item When the fermions are far from the Higgs:
	\begin{eqnarray}
	\frac{X}{\rho_\psi}\frac{\rho_H^2}{\rho_\psi^2}\gg 1 &\rightarrow& \lambda = \frac{g}{\rho_{\psi\text{phys}}^2}\left( 	
		\frac{\rho_{\psi}}{X}\right)^4\nn\\
	\frac{X}{\rho_\psi}\frac{\rho_H^2}{\rho_\psi^2}\ll 1 &\rightarrow& \lambda = \frac{g}{\rho_{\psi\text{phys}}^2}\left(	
		\frac{\rho_{\psi}}{X}\right)^3 e^{-2\kappa (Y_\psi-Y_H)}\nn
	\end{eqnarray}
\end{itemize}
 
\


\begin{table}[ht]
\begin{center}
\begin{tabular}{|c|cl||}
\hline&&\\
limits & & $g \int d^4z \, \tilde{\psi}_i^\dagger(z)  \Phi_H(z) \tilde{\psi}_j(z)$ \\ && \\ \hline&&\\
$\rho_H\sim\rho_\psi\ll X$ &   $\frac{g}{\rho_{H}^2}$&$\left( \frac{\rho_H}{X}\right)^3$  \\ && \\
$\rho_H\ll\rho_\psi\sim X$ &  $\frac{g}{\rho_{H}^2}$ & $\left( \frac{\rho_H}{X}\right)^2$   \\ &&\\
$\rho_H\ll\rho_\psi\ll X$ &  $\frac{g}{\rho_{H}^2}$ & $\left( \frac{\rho_H}{X}\right)^2 \left( \frac{\rho_\psi}{X}\right)^2 
				\left[ 1+\frac{X}{\rho_\psi}\left(\frac{\rho_H}{\rho_\psi}\right)^2\right]$  \\ &&\\
$\rho_H\ll X \ll \rho_\psi$ &  $\frac{g}{\rho_{H}^2}$ & $\left( \frac{\rho_H}{\rho_\psi}\right)^4  
				\left[ 1+\left(\frac{X}{\rho_H}\right)^2\left(\frac{X}{\rho_\psi}\right)^2\right]$  \\ &&\\
$X \lesssim \rho_H \ll \rho_\psi$ &  $\frac{g}{\rho_{H}^2}$ & $ \left( \frac{\rho_H}{\rho_\psi}\right)^4$  \\ &&\\\hline
\hline
\end{tabular}\caption{Various limits of the integral giving the Yukawa coupling. \label{tab}}
\end{center}
\end{table}

\section{Comments}

We have seen that there is a rather intricate string theory picture underlying many of the important
features of the 5d warped phenomenology models. One might ask: can we distinguish the string theory models
from the 5d phenomenology?

Obviously, yes in principle: the spectrum of the 5d models consists of the zero modes
which become the standard model particles after symmetry breaking; then in addition, 
for each standard model particle there is an infinite Kaluza-Klein tower of resonances 
with the {\it same spin} as its associated
standard model cousin. These particles are also present in the string spectrum, but the string
theory has more: for each standard model particle, there is also an {\it infinite tower of string states
of increasing spins}. So, measuring even part of the spectrum could be enough to distinguish them.%
\footnote{Usually, in the holographic limit \cite{malda} we decouple these massive open string states, but
here we cannot since the string length and string coupling is finite.}

In the 5d models, the masses of the Kaluza-Klein modes are typically quantised in units of a TeV.
Therefore, the LHC will only be sensitive to the first or second resonance. What about the string states?
The $AdS_5$ scale is of order $m_p$ so, for weak string coupling the string scale is below this.
However, the $D$7-branes fill the entire $AdS_5$ and hence, the 7-7 strings which are in the
infrared end of $AdS_5$ will have a TeV scale or lower mass: 
hence only the first or second of these will be directly
accessible at the LHC. Since these states have the same gauge quantum numbers as the Kaluza-Klein
modes, they could only be distinguished by their decay patterns or their spins. For example, there
might be a spin 3/2 colored particle which is a string excitation of the gluon. If produced,
this particle must eventually decay into jets, the angular distributions of which will be sensitive
to its spin. It is interesting to study to what extent these events can be selected
and the discovery reach for the LHC \cite{wip}.

\

We conclude with a discussion of some additional issues which deserve further investigation.
Firstly there is the issue of supersymmetry breaking. In 5d models, one does not a priori need
supersymmetry at all, since the electroweak scale is generated through the warped extra dimension.
But in Type IIB string theory, there is certainly local supersymmetry in the UV, and one needs to
break it. One possibility is to choose the background fluxes and geometry to explicitly break supersymmetry,
such as was recently considered in \cite{giddingsquevedo}. 
However, backgrounds which explicitly break supersymmetry in string theory can often be unstable; thus,
it would be good to investigate this further.
Secondly, there is the issue of fermion chirality. With one collection of parallel $D$7-branes,
even though the backgrounds we have considered generate multiple copies of the same standard model
representations, the representations include both fermion chiralities. This can be avoided by
the introduction of another set of $D$7-branes intersecting the first set along a surface in $\Sigma$,
but we have not investigated this in detail. Also, in 5d models, the chirality problem is resolved by
considering a $\mathbb{Z}_2$ orbifold and perhaps such a mechanism can also be realised in string theory.

\vskip 2cm

\begin{center} \textbf{Acknowledgments} \end{center}

We would like to thank Joseph Conlon, Paolo Creminelli, Michael R. Douglas, Johanna Erdmenger, Shamit Kachru, Maurizio Piai and Marco Serone for useful and stimulating discussions.

\newpage

\appendix

\section{The ADHM Construction}

In this section we briefly review the ADHM formalism for instantons and how to use it to find bosonic and fermionic zero modes around their background \cite{Atiyah:1978ri} (to have a more complete review of the subject, see \cite{dorey} and references therein). We are interested in constructing finite action solutions of the four dimensional Euclidean Yang-Mills theory (instantons). The gauge potential satisfies a first order (anti-)self-duality equation
\begin{equation} \label{01}
F_{\mu\nu} = \pm (\ast F)_{\mu\nu} = \pm \frac{1}{2} \epsilon_{\mu\nu\rho\sigma} F_{\rho\sigma}
\end{equation}
In the following we will restrict ourselves to $U(N)$ gauge groups.

In order to discuss the ADHM formalism, we introduce the quaternionic notation:
\begin{gather}
\mathbf{z} = z_\mu \sigma_\mu \qquad \qquad \bar{\mathbf{z}} = z_\mu \bar\sigma_\mu \\
z_\mu = \frac{1}{2} \tr \:\mathbf{z} \:\bar\sigma_\mu
\end{gather}
where $\sigma_\mu=(i\tau^a,1)$ and $\bar\sigma_\mu=(-i\tau^a,1)$. 

The ADHM formalism allows to obtain (anti-)self-dual field strength configurations by solving only algebraic equations.
The gauge field with instanton number $k$ for $U(N)$ gauge group is given by
\begin{equation}
A_\mu = v(z)^\dag \partial_\mu v(z) \:,
\end{equation}
where $v(z)$ is a $(N+2k)\times N$ matrix. It is defined by the equations
\begin{gather}
v(z)^\dag v(z) = 1 \label{02} \\
v(z)^\dag \Delta(z) = 0 \:. \label{03}
\end{gather}
Here $\Delta(z)$ is a $(N+2k)\times 2k$ matrix, linear in the position variable $z$, having the structure
\begin{equation} \label{04}
\Delta(z) = \begin{cases}
a - b \mathbf{z} & \text{self-dual instantons,} \\
a - b \bar{\mathbf{z}} & \text{anti-self-dual instantons,} \end{cases}
\end{equation} 
The matrices $a,b$ are constrained to satisfy the condition
\begin{equation}\label{eqR003}
\Delta(z)^\dag \Delta(z)=p^{-1}(z) \otimes \mathbbm{1}_2
\end{equation}
where $p^{-1}(z)$ is a $k\times k$ invertible matrix. This assures the (anti-)self-duality equation \eqref{01}. 

$a,b$ are $(N+2k)\times 2k$ matrices that contain the moduli of the instantonic configuration. 
Beacause of some symmetries of the equations above they can be brought to the form
\begin{equation}
a = \begin{pmatrix} \lambda \\ \xi \end{pmatrix} \qquad \qquad b = \begin{pmatrix} 0 \\ \mathbbm{1}_{2k} \end{pmatrix} \:,
\end{equation}
where $\lambda$ is an $N\times 2k$ matrix and $\xi$ is a $2k\times 2k$. There is no one-to-one correspondence between these two matrices and the moduli: some constraints and redundancies are left. The actual number of moduli is $4Nk$.

\subsubsection*{Fermion Zero Modes}

We will be interested in the fermionic zero modes in the fundamental representation and with definite chirality, {\it i.e.} those solving:
\begin{equation}
 \sigma^\mu D_\mu \eta = \sigma^\mu (\partial_\mu+v^\dagger \partial_\mu v)\eta \:.
\end{equation}

One gets $k$ independent solutions for $\eta^T$ as an $N\times 2$ matrix:
\begin{equation}\label{eqR006}
 \eta^i_{u,\alpha}=(v^\dagger b p \sigma^2)_{u,i\alpha}
\end{equation}
where $u=1,...,N$, $i=1,...,k$ and $\alpha=1,2$. Thus we have found $k$ fermionic zero modes in the fundamental representation.

\subsection{$k=1$ $SU(2)$ Instanton}

We apply the machinery described above to the simplest case of one $SU(2)$ instanton. In this case, applying the further constraints on $a$ and $b$, one can put $\Delta$ in the form:
\begin{equation}
 \Delta = \left(\begin{array}{c} \rho \, \mathbf{1}_2 \\ \bar{\mathbf{Z}}-\bar{\mathbf{z}}\end{array}\right)
\end{equation}
with $\rho$ and $Z_\mu$ the $(4kN - N^2+1) = 5$ parameters of the solution in the case $k=1,N=2$. 

From here, using \eqref{eqR003}, we can get $f$:
\begin{equation}
 p(z) = \frac{1}{\rho^2+(z-Z)^2}
\end{equation}
Then solving for the normalized zero eigenvectors $v^\dagger \Delta=0$ and $v^\dagger v=1$, we have:
\begin{equation}
 v(z) = \left(\begin{array}{c} \left(\frac{(z-Z)^2}{\rho^2+(z-Z)^2}\right)^{1/2}\mathbf{1}_2   \\
             \left(\frac{\rho^2}{(z-Z)^2(\rho^2+(z-Z)^2}\right)^{1/2} (\mathbf{z-Z}) \end{array}\right)
\end{equation} 
And finally one gets the connection in singular gauge:
\begin{equation}
  A_\mu= \frac{\rho^2 (z-Z)_\nu}{(z-Z)^2(\rho^2+(z-Z)^2}\sigma_{\mu\nu}
\end{equation}

\subsubsection*{Fermion Zero Modes}

We compute the fermion zero modes in this simple anti-instanton background, by using the formula \eqref{eqR006}:
\begin{equation}
  v^\dagger b p = \frac{\rho}{(\rho^2+(z-Z)^2)^{3/2}} \frac{\mathbf{z-Z}}{|z-Z|}
\end{equation} 
This is a $2\times 2$ matrix. One index is for the fundamental rep, while the other is a spinorial index.

\subsection{'t Hooft Solution}

Now we consider the case in which $k$ is general, the gauge group is $SU(2)$, 
and we will concentrate on a class of solutions described by $5k$ parameters (instead of $8k$): $\rho_i$ and $Z_i$, with $i=1,...,k$. It is called the {\it 't Hooft solution} \cite{tHooft} and is characterized, in the ADHM construction, by:
\begin{equation}
 v(z)= \left(\begin{array}{c} \left[1+\sum_{i=1}^k\frac{\rho_i^2}{(z-Z_i)^2}\right]^{-1/2}\mathbf{1}_2   \\
             \left[1+\sum_{i=1}^k\frac{\rho_i^2}{(z-Z_i)^2}\right]^{-1/2}\frac{\rho_i^2(\mathbf{z-Z}_i)}{(z-Z_i)^2}\end{array}\right)
\end{equation}
It is obtained by taking
\begin{equation}
a = \begin{pmatrix} \rho_i\mathbf{1}_2 \\ \delta_{ji}\mathbf{Z}_i \end{pmatrix} \qquad \qquad 
b = \begin{pmatrix} 0 \\ \mathbbm{1}_{2k} \end{pmatrix} \:,
\end{equation}
From these, one can also get the expression for $p$. The diagonal entries are:
\begin{equation}
 p_{ii} = \left[1+\sum_{\ell=1}^k\frac{\rho_\ell^2}{(z-Z_\ell)^2}\right]^{-1}\frac{1}{(z-Z_i)^2}
	\left[1+\sum_{j\not=i}\frac{\rho_j^2}{(z-Z_j)^2}\right]\:,
\end{equation}
while the off-diagonal elements are:
\begin{equation}
 p_{ij}=-\left[1+\sum_{\ell=1}^k\frac{\rho_\ell^2}{(z-Z_\ell)^2}\right]^{-1} \frac{\rho_i\rho_j}{(z-Z_i)^2(z-Z_j)^2}\:.
\end{equation}

\

There are asymptotic regions of the parameters space where the multi-instanton configurations can be identified as being composed of well-separated single instantons. One can show that this limit is valid when
\begin{equation}\label{eqR012}
 (Z_i-Z_j)^2\gg \rho_i\rho_j \qquad \forall i\not = j
\end{equation}
In this limit the $Z_i$'s become the positions of the $k$ instantons, while the $\rho_i$'s are their sizes.

\subsubsection*{Fermion Zero Modes}
As in the case $k=1$, we compute the fermion zero modes in the background described above, by using the formula \eqref{eqR006}:
\begin{eqnarray}\label{eqR014}
  (v^\dagger b p)_h &=& \left[1+\sum_{\ell=1}^k\frac{\rho_\ell^2}{(z-Z_\ell)^2}\right]^{-3/2}\frac{\rho_h}{(z-Z_h)^2}\times\\
	&&\times  \left\lbrace\left[1+\sum_{\ell=1}^k\frac{\rho_\ell^2}{(z-Z_\ell)^2}\right]\frac{\mathbf{z-Z}_h}{(z-Z_h)^2}-
		 \sum_{j=1}^k\frac{\rho_j^2}{(z-Z_j)^4}(\mathbf{z-Z}_j)\right\rbrace \nn
\end{eqnarray}
It is in the fundamental representation of $SU(2)$.

\

In the limit of well separated $k$ instantons, {\it i.e.} \eqref{eqR012}, the expression for the fermionic zero modes simplifies:
\begin{eqnarray}\label{eqR018}
  (v^\dagger b p)_h &\sim& \frac{\rho_h}{(\rho_h^2+(z-Z_h)^2)^{3/2}} \frac{\mathbf{z-Z}_h}{|z-Z_h|}\:.
\end{eqnarray}
It is the same expression for the fermion zero mode in the case of one instanton localised in $Z_h$. One can see that in regions around other instanton ($z\sim Z_j$, with $j\not =h$), the solution found above is of order $\frac{\rho_h \rho_j}{(Z_h-Z_j)^2}\ll 1$. So in this approximation there is one fermionic zero mode localised around each instanton. One has to note that the suppression of points distant from every instanton positions is larger than that obtained  around $Z_{j\not=k}$. On these points we have low peak, suppressed with respect to that on $Z_h$, but larger with respect to the value of the single instanton profile at that point.\label{page}

\subsubsection*{Vector Zero Modes}

The vector zero modes in the adjoint representation of $SU(2)$ are those variations of $A_\mu$ that leave it a solution of the (anti-)selfdual equation (and that are not gauge transformations). They are associated to the parameter that describe the solution. 

In the ADHM construction it is given the expression of the zero modes:
\begin{equation}
 \delta A_\mu = -v^\dagger(\delta \! a p \sigma_\mu b^\dagger - b\bar{\sigma}_\mu p \delta\! a^\dagger)v
\end{equation}

Consider again the 't Hooft solution. There are $5k$ zero modes: $4k$ associated with changing positions of each instanton, and $k$ with changing their sizes. 

As an example, we give the expression for the zero mode relative to the translation of $Z_j$, by the vector $\Phi$:
\begin{eqnarray}\label{eqR015}
 \delta A^{\Phi}_\mu &=& \Phi_\nu\left[1+\sum_{\ell=1}^k\frac{\rho_\ell^2}{(z-Z_\ell)^2}\right]^{-2} \frac{\rho_j^2}{(z-Z_j)^4}
 	(\mathbf{z-Z}_j)^\dagger \sigma_{\mu\nu}\,\times \nn\\
		&&\times\,\left\lbrace \frac{(\mathbf{z-Z}_j)}{(z-Z_j)^2}\left[1+\sum_{i\not=j}\frac{\rho_i^2}{(z-Z_i)^2}\right]
	- \sum_{i\not=j}\frac{\rho_i^2}{(z-Z_i)^4}(\mathbf{z-Z}_i)\right\rbrace \nn\\
\end{eqnarray}
One can see that in the limit \eqref{eqR012} it becomes the zero mode of the single instanton solution localised on $Z_j$:
\begin{equation}
 \delta A^{\Phi}_\mu \sim \Phi_\nu\frac{\rho_j^2}{(z-Z_j)^2}
 	\frac{(\mathbf{z-Z}_j)^\dagger \sigma_{\mu\nu} (\mathbf{z-Z}_j)}{(\rho_j^2+(z-Z_j)^2)^2}
\end{equation}

\newpage

\section{Warping Effects on the Dirac Operator}

We want to find the spin-connection relative to the metric:
\begin{eqnarray}
 ds^2 &=& f(r)^{-1/2}\tilde{g}_{(3,1)\mu\nu}\,dx^\mu dx^\nu + f(r)^{1/2} \tilde{g}_{(4)\alpha\beta}\,dz^\alpha dz^\beta \nn\\
      &=& f(r)^{-1/2}\eta_{mn} \, \tilde{e}^m\tilde{e}^n + f(r)^{1/2} \delta_{ab} \, \tilde{e}^a\tilde{e}^b\\
      &=& \eta_{mn}\, e^m e^n + \delta_{ab}\, e^a e^b \nn = \eta_{HK}\, e^H e^K
\end{eqnarray}
The corresponding 8-bein is given then by $e^m=f(r)^{-1/4}\tilde{e}^m$ and $e^a=f(r)^{1/4}\tilde{e}^a$. $r$ is the radial coordinate in the 4-dimensional space spanned by the coordinates $z^\alpha$.

The spin connection is given by:
\begin{eqnarray}
 \omega_\Pi^{HK}&=&\frac{1}{2}e^{\Lambda H}(\partial_\Pi e^K_\Lambda-\partial_\Lambda e^K_\Pi) - [H\leftrightarrow K]\nn\\
 	&& -\frac{1}{2}e^{\Xi H}e^{\Upsilon K} (\partial_\Xi e_{\Upsilon Q}-\partial_\Upsilon e_{\Xi Q})e_\Pi^Q
\end{eqnarray}
Using this formula, one obtains:
\begin{eqnarray}
 \omega_\mu^{ab} &=& \tilde{\omega}_\mu^{ab} \nn\\
 \omega_\mu^{an} &=& \tilde{\omega}_\mu^{an}+\frac{f'}{4f^{3/2}}\tilde{e}_\mu^n \tilde{e}^{ra}\nn\\
 \omega_{\beta}^{\hat{a}\hat{b}} &=& \tilde{\omega}_{\beta}^{\hat{a}\hat{b}}\nn\\
 \omega_{\hat{\beta}}^{\hat{a}R} &=& \tilde{\omega}_{\hat{\beta}}^{\hat{a}R}
 		+\frac{f'}{4f} \tilde{e}_{\hat{\beta}}^{\hat{a}}\nn\\
 \omega_{r}^{\hat{a}R} &=& \tilde{\omega}_{r}^{\hat{a}R} \nn\\		
\end{eqnarray}
where $\tilde{\omega}$ is the spin connection associated to $\tilde{g}$ and the coordinates $x^\alpha$ are split in the radial coordinate $r$ and in the other three coordinates $x^{\hat{\alpha}}$ (and also $a=R,\hat{a}$).

\

The Dirac operator is given by
\begin{equation}
 \D_8 = e^\Pi_K \Gamma^K (\partial_\Pi + \omega_\Pi^{HQ} \frac{1}{4}\Gamma_H\Gamma_Q + A_\Pi)
\end{equation}
In the setup we are considering ($A_\mu=0$ and $\omega$ given above), it is equal to
\begin{eqnarray}
\D_8 &=& f^{1/4} \tilde{e}^\mu_m \Gamma^m (\partial_\mu + \omega_\mu^{HQ} \frac{1}{4}\Gamma_H\Gamma_Q)+
f^{-1/4} \tilde{e}^\alpha_a \Gamma^a (\partial_\alpha + \omega_\alpha^{HQ} \frac{1}{4}\Gamma_H\Gamma_Q + A_\alpha)\nn\\
&=& f^{1/4} \tilde{e}^\mu_m \Gamma^m ((\tilde{D}_{3,1})_\mu + \delta\omega_\mu^{HQ} \frac{1}{4}\Gamma_H\Gamma_Q)+
f^{-1/4} \tilde{e}^\alpha_a \Gamma^a ((\tilde{D}_4)_\alpha + \delta\omega_\alpha^{HQ} \frac{1}{4}\Gamma_H\Gamma_Q)\nn
\end{eqnarray}
Where $\delta\omega=\omega-\tilde{\omega}$ can be read off above. In particular 
\begin{eqnarray}
\tilde{e}^\mu_m \Gamma^m \, \delta\omega_\mu^{HQ} \frac{1}{4}\Gamma_H\Gamma_Q &=& 
\frac{3}{8}\frac{f'}{f^{3/2}} \Gamma_r \\
\tilde{e}^\alpha_a \Gamma^a \,  \delta\omega_\alpha^{HQ} \frac{1}{4}\Gamma_H\Gamma_Q &=& 
-\frac{f'}{2f} \Gamma_r
\end{eqnarray}
Putting all together one gets:
\begin{eqnarray}
\D_8 &=& f^{1/4} (\tilde{\D}_{3,1}\otimes {\bf 1}   +\frac{3}{8}\frac{f'}{f^{3/2}}\gamma^{(4)}\otimes\gamma_r) + f^{-1/4} (\gamma^{(4)}\otimes\D_4 -\frac{f'}{2f}\gamma^{(4)}\otimes\gamma_r)\nn\\
&=& f^{1/4} \tilde{\D}_{3,1}\otimes {\bf 1}+ f^{-1/4}\gamma^{(4)}\otimes\tilde{\D}_4 - \frac{1}{8f^{1/4}}\frac{f'}{f} \gamma^{(4)}\otimes\gamma_r
\end{eqnarray}

\

Splitting the 8-dimensional spinor as $\Psi = \sum_k \chi_k(x) \otimes \psi_k(y) $, we see that the zero modes of $\tilde{\D}_{3,1}$ are associated to the zero modes of the operator $\hat{\D}_4=\tilde{\D}_4 - \frac{f'}{8f} \gamma_r$. If $\psi_0$ is a zero mode of $\tilde{\D}_4$, then $\psi=f^{1/8}\psi_0$ is a zero mode of $\hat{\D}_4$, since:
\begin{eqnarray}
\tilde{\D}_4 (f^{1/8}\psi_0) = \gamma_r (\partial_r f^{1/8}) \psi_0 = \frac{f'}{8f}\gamma_r(f^{1/8} \psi_0)\:.
\end{eqnarray}

\end{document}